\documentclass[twocolumn]{aastex62}
\hypersetup{linkcolor=magenta,citecolor=blue,filecolor=cyan,urlcolor=purple}

\usepackage{graphicx}
\usepackage[caption=false]{subfig}
\usepackage{ulem}

\received{2019 October 2}
\revised{2020 February 20}
\accepted{2020 March 11}
\submitjournal{ApJ}

\shorttitle{Evolution of ICME Sheath Fluctuations}
\shortauthors{Good et al.}

\begin{document}

\title{Radial Evolution of Magnetic Field Fluctuations in an Interplanetary Coronal Mass Ejection Sheath}

\correspondingauthor{S. W. Good}
\email{simon.good@helsinki.fi}

\author[0000-0002-4921-4208]{S. W. Good}
\affil{Department of Physics, University of Helsinki, Helsinki, Finland}

\author[0000-0001-9574-339X]{M. Ala-Lahti}
\affil{Department of Physics, University of Helsinki, Helsinki, Finland}

\author[0000-0001-6590-3479]{E. Palmerio}
\affil{Department of Physics, University of Helsinki, Helsinki, Finland}
\affil{Space Sciences Laboratory, University of California–-Berkeley, Berkeley, USA}

\author[0000-0002-4489-8073]{E. K. J. Kilpua}
\affil{Department of Physics, University of Helsinki, Helsinki, Finland}

\author[0000-0003-2555-5953]{A. Osmane}
\affil{Department of Physics, University of Helsinki, Helsinki, Finland}



\begin{abstract}

The sheaths of compressed solar wind that precede interplanetary coronal mass ejections (ICMEs) commonly display large-amplitude magnetic field fluctuations. As ICMEs propagate radially from the Sun, the properties of these fluctuations may evolve significantly. We have analyzed magnetic field fluctuations in an ICME sheath observed by \textit{MESSENGER} at 0.47~au and subsequently by \textit{STEREO-B} at 1.08~au while the spacecraft were close to radial alignment. Radial changes in fluctuation amplitude, compressibility, inertial-range spectral slope, permutation entropy, Jensen-Shannon complexity, and planar structuring are characterized. These changes are discussed in relation to the evolving turbulent properties of the upstream solar wind, the shock bounding the front of the sheath changing from a quasi-parallel to quasi-perpendicular geometry, and the development of complex structures in the sheath plasma.

\end{abstract}



\section{Introduction}

The solar wind magnetic field is characterized by fluctuations across a broad range of timescales. The \textit{k}-space power spectrum of these fluctuations typically displays a spectral index near --5/3 in the inertial range, consistent with a hydrodynamic Kolmogorov turbulent cascade of energy from larger to smaller timescales. This intermediate range is observed at spacecraft frame frequencies $f\lesssim 10^{-1}$~Hz at 1~au. At higher frequencies, dissipative kinetic processes tend to dominate and the spectral slope steepens. A third spectral range with index --1 is observed in fast solar wind, at $f\lesssim 10^{-3}$~Hz at 1~au. The $1/f$ range is commonly attributed to large-amplitude Alfv\'en waves propagating from the solar corona \citep[e.g.,][]{Velli89}; an alternative explanation has recently been proposed by \citet{Matteini18}, who show how a $1/f$ spectral range naturally arises when the amplitudes of Alfv\'enic fluctuations reach a limit imposed by their incompressibility, regardless of the fluctuation origin.

The evolution of magnetic field fluctuations in the solar wind with distance from the Sun has been extensively studied. The spectral break point between the $1/f$ and inertial ranges in fast wind moves to lower frequencies with distance \citep{Bavassano82}, taken to be evidence of a locally active turbulent cascade. Fluctuations become progressively less Alfv\'enic with increasing frequency in the inertial range and generally less Alfv\'enic across the inertial range with heliocentric distance \citep[e.g.,][]{Marsch90}. Most radial evolution studies have considered statistically averaged properties, while a relatively small number have analyzed the same fast solar wind intervals observed by radially aligned spacecraft \citep{Schwartz83,D'Amicis10,Bruno14,Telloni15}. Opportunities to perform such studies have been rare given the scarcity of spacecraft alignments. \citet{Schwartz83} discuss the value and limitations of line-up studies of individual plasma parcels versus statistical studies of plasma parameters averaged at different radial distances.

Among the most strongly fluctuating and turbulent of space plasma environments are the sheaths of interplanetary coronal mass ejections (ICMEs). ICME sheaths consist of piled-up solar wind preceding relatively fast ICMEs propagating away from the Sun. When the speed difference between an ICME and the ambient solar wind exceeds the local fast magnetosonic wave speed, a fast forward shock will form at the sheath leading edge. An ICME sheath is in many respects the solar-transient equivalent of a planetary magnetosheath. However, ICME sheaths have lower Mach number shocks with much greater spatial extensions than their planetary counterparts, and there tends to be a much weaker non-radial deflection (and hence more accumulation) of plasma within ICME sheaths \citep{Siscoe08}. ICME sheaths also share some general properties of the fast solar wind (e.g., high fluctuation amplitudes) and slow wind (e.g., lower Alfv\'enicity). Plasma fluctuations found in ICME sheaths may comprise pre-existing fluctuations from the swept-up solar wind and fluctuations generated locally within the sheath, and may include a broad range of wave activity \citep[e.g.,][]{Liu06,Ala-Lahti18,Ala-Lahti19}. Global properties of the sheaths such as bulk flow speed can modify sheath turbulence properties, particularly in the kinetic range \citep{Riazantseva19}. A comprehensive review of ICME sheath properties, including their significant space weather impact, is provided by \citet{Kilpua17}.

In this paper, we present the first study of how magnetic field fluctuations evolve in an ICME sheath observed at two radially aligned spacecraft, and focus on the turbulent nature of the fluctuations in the inertial range. The observing spacecraft, \textit{MESSENGER} and \textit{STEREO-B}, were located at radial distances of 0.47~au and 1.08~au, respectively. Changes in fluctuation amplitude, fluctuation amplitude normalized to the mean field, and fluctuation compressibility as functions of timescale have been determined. The values of these parameters in the sheath plasma are also compared to their values in the solar wind immediately ahead of the sheath. Although the scope of the investigation is somewhat limited by the absence of solar wind plasma measurements at \textit{MESSENGER}, much can still be determined from the magnetic field data alone.

We also apply recently developed analysis techniques to quantify the permutation entropy and Jensen-Shannon complexity \citep{Bandt02,Rosso07} of the sheath fluctuations at each spacecraft. These two quantities can indicate whether the physical processes that generate the fluctuations are fundamentally stochastic or chaotic, and can indicate the relative abundance of coherent structures versus stochastic fluctuations. Previous studies have found solar wind fluctuations to be highly stochastic \citep{Weck15,Olivier19}, consistent with them having a largely turbulent origin. \citet{Weygand19} found a general increase in entropy and decrease in complexity in turbulent solar wind intervals with heliocentric distance, possibly due to evolution in the turbulent cascade. In their analysis of fractal dimensions, \citet{Munoz18} found magnetic field time series from ICME sheaths to be highly complex when compared to time series from more typical solar wind and the ICME flux rope drivers.

We note here an important contextual difference between previous radial alignment studies of the solar wind and the present study. Previous studies have sought to capture the unperturbed evolution of steady-source solar wind streams observed away from stream interaction regions, ICMEs, and other potential sources of fluctuations generated locally in interplanetary space. These studies have thus examined the progressive `aging' with radial distance of fluctuations that primarily arise from energy injected at the Sun. In contrast, we study the evolution of a propagating interaction region in which local sources of fluctuations (e.g., the shock bounding the sheath) may be significant and fluctuations may be relatively `young' in age. We emphasize that the launch and propagation of ICMEs are generally associated with a range of fluctuation-generating heliospheric activity: for example, a large fraction of the heliosphere can be influenced by solar energetic particles (SEPs) accelerated by ICME shocks, and the interplanetary magnetic field may be globally reordered by the passage of the coherent, large-scale ICME structure.

\section{Spacecraft Observations}

\begin{figure*}
\epsscale{1}
\plotone{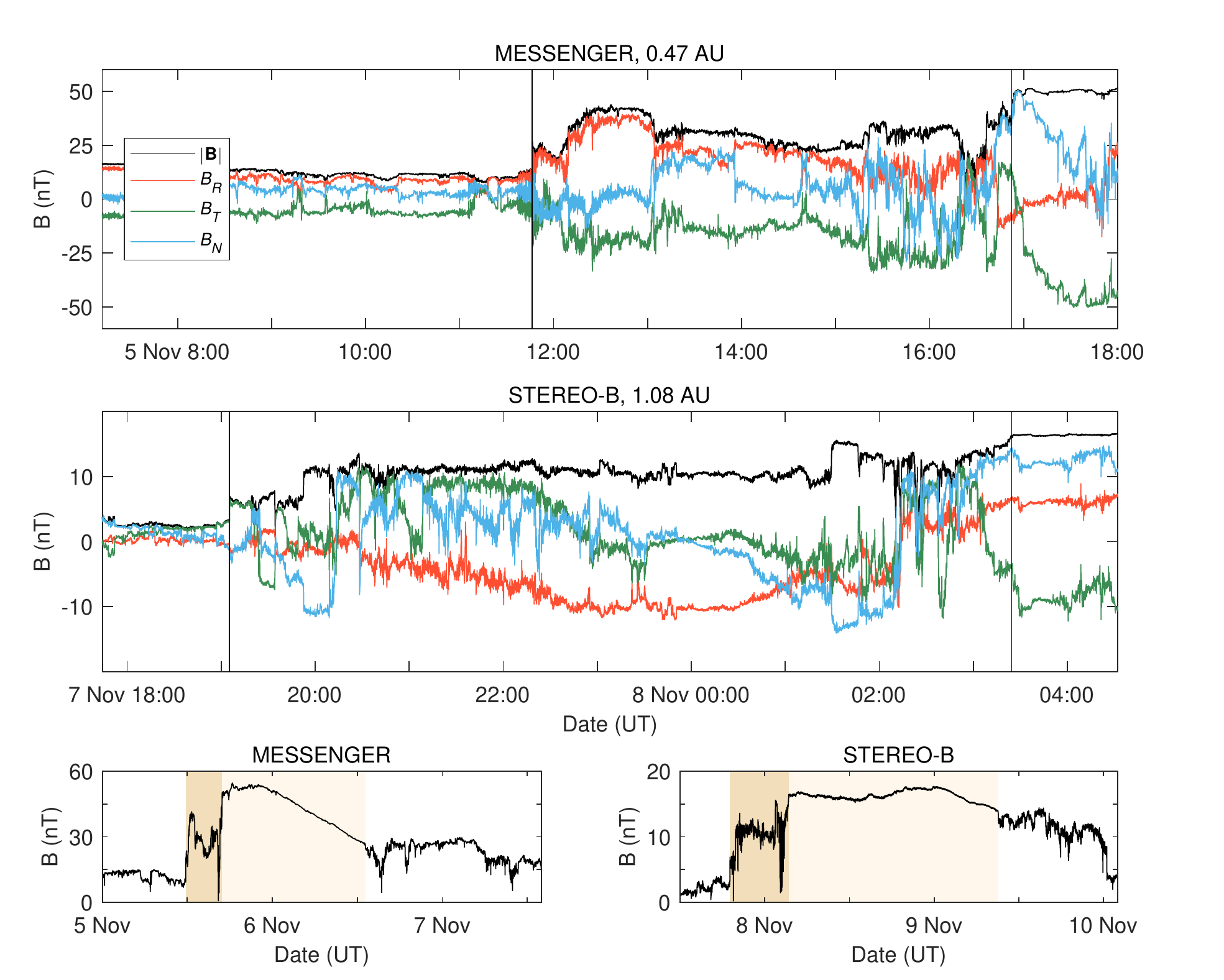}
\caption{Magnetic field observations in RTN coordinates at \textit{MESSENGER} and \textit{STEREO-B}. Black vertical lines demarcate the sheath interval in the top and middle panels. In the bottom left and right panels, which show the field magnitude over a longer time period, the ICME sheath and flux rope are shaded in the darker and paler shades of beige, respectively.}
\label{fig:B_data}
\end{figure*}

Magnetic field data with a time resolution of 0.5~s from \textit{MESSENGER} \citep{Anderson07} and 0.125~s from \textit{STEREO-B} \citep{Acuna08} are analyzed in this study. Figure~\ref{fig:B_data} displays the magnetic field observations during the sheath passage at each spacecraft. The ICME-driven shock arrived at \textit{MESSENGER} at 2010 Nov 7 11:46~UT. Approximately 55~hr later, at 2010 Nov 7 19:05~UT, the shock arrived at \textit{STEREO-B}. The two spacecraft were separated by 0.61~au in radial distance, 1$^{\circ}$ in heliographic longitude and 7$^{\circ}$ in heliographic latitude around this time. At both spacecraft, the sheath was bounded to the rear by the leading edge of the ICME's flux rope. The sheath duration was 5~hr 6~minutes at \textit{MESSENGER} and 8~hr 20~minutes at \textit{STEREO-B}. Wave activity in the sheath at \textit{STEREO-B} has recently been analyzed by \citet{Li19}. The sheath at \textit{STEREO-B} was preceded by a gradual SEP event, as indicated by a rise in the 1.8--3.6~MeV proton intensity detected at the spacecraft from around Nov 4 07:30~UT, a sharper rise in intensity at Nov 5 02:30~UT, and a shock-spike enhancement at Nov 5 20:00~UT.

\subsection{Shock Orientations}

The IP Shocks Database (www.ipshocks.fi), which provides interplanetary shock normals derived with the mixed--mode method \citep[`MD3' method in][]{Abraham-Shrauner76}, lists the leading shock normal at \textit{STEREO-B} as [0.97 0.22 -0.05] in RTN coordinates. The shock was quasi-perpendicular at 1.08~au, with a shock normal oriented at $85^{\circ}$ relative to the upstream magnetic field. 

The mixed--mode method requires velocity measurements as input and so cannot be applied to shocks that were observed by \textit{MESSENGER}. We have therefore used the magnetic coplanarity theorem \citep{Colburn66}, which only requires magnetic field observations, to estimate the shock normal at \textit{MESSENGER}. The \textit{STEREO-B} shock normal has also been estimated with this method and compared to the IP Shocks listing. In the coplanarity theorem, the shock normal $\hat{\textbf{n}}$ is related to the upstream ($\textbf{B}_u$) and downstream ($\textbf{B}_d$) magnetic fields by

\[
\hat{\textbf{n}}=\pm \frac{(\textbf{B}_u \times \textbf{B}_d) \times (\textbf{B}_u - \textbf{B}_d)}
    {|(\textbf{B}_u \times \textbf{B}_d) \times (\textbf{B}_u - \textbf{B}_d)|}.
\]

Following the IP Shocks methodology, the upstream and downstream field vectors were averaged over 8-minute intervals ending 1 minute before and starting 2 minutes after the shock time to give $\textbf{B}_u$ and $\textbf{B}_d$, respectively. The 3-minute cut-out was made to remove strong wave-like fluctuations in the vicinity of the shock interface. In RTN coordinates, the anti-sunward shock normals were found to be [0.57 -0.24 0.79] at \textit{MESSENGER} and [0.21 0.40 0.89] at \textit{STEREO-B}. These normals were oriented $34^{\circ}$ and $68^{\circ}$ relative to the upstream magnetic field direction at \textit{MESSENGER} and \textit{STEREO-B}, respectively, suggesting the shock was quasi-parallel at the inner spacecraft and quasi-perpendicular at the outer spacecraft. Evolution from a quasi-parallel to perpendicular shock geometry with heliocentric distance is to be expected generally, given the form of the Parker spiral.

Although the shock-to-upstream field angles estimated with the two methods are similar at \textit{STEREO-B}, the shock directions themselves differ significantly. The limitations of the coplanarity theorem have been noted previously \citep{Schwartz98}, and the values that we have derived from it should be treated with some caution. The IP Shocks listing for the shock at \textit{STEREO-B} is taken to be the more robust orientation estimate.

\subsection{Taylor's Hypothesis}

In the following analysis, we consider fluctuations in the spacecraft frame, and compare fluctuations in the sheaths observed at each spacecraft with fluctuations in the upstream solar wind. It is assumed that Taylor's hypothesis, which states that a spacecraft trajectory through plasma represents an instantaneous spatial cut when the dynamical timescales of the fluctuations are much less than the advection timescale, is valid. This assumption allows fluctuation timescales $\Delta t$ to be straightforwardly related to fluctuation length scales $l=v \Delta t$, where $v$ is the flow speed and $l$ is the length scale sampled in the flow direction. One simple measure of Taylor hypothesis validity in a particular plasma environment is to check whether $v_A/v \ll 1$ is satisfied, where $v_A$ is the Alfv\'en speed. This criterion is valid for Alfv\'en waves in the inertial range, and assumes that wavevectors perpendicular to the mean magnetic field are much greater than parallel wavevectors \citep{Howes14}. In the super-Alfv\'enic solar wind (where Taylor's hypothesis is generally very well satisfied) upstream of the sheath at \textit{STEREO-B}, $v_A/v = 0.06$, comparable to the values in the sheath, where $v_A/v = 0.19$. The $v_A/v$ ratio was unknown at \textit{MESSENGER} due to the lack of plasma measurements, but assuming constant plasma flow speeds and an inverse-square fall-off in density from 0.47~au to the values measured at 1.08~au gives $v_A/v$ estimates of 0.14 in the solar wind and 0.22 in the sheath at \textit{MESSENGER}. 

There is also some Doppler effect associated with the solar wind-sheath transition in the spacecraft frame such that fluctuation timescale $\Delta t$ in the solar wind is shifted to $\Delta t'=(v/v')\Delta t$ in the sheath, where $v$ and $v'$ are the solar wind and sheath flow speeds, respectively. At \textit{STEREO-B}, this shift was relatively small ($v/v'\sim 0.83$) when compared to the logarithmic scalings of $\Delta t$ that are investigated, and so has been neglected; it is assumed that $v/v'$ was of a similar order of magnitude at \textit{MESSENGER}, and the Doppler shift there has likewise been neglected. 

\smallskip

\section{Magnetic Fluctuations}

We have determined magnetic field fluctuations $\delta\textbf{B}(t,\Delta t)=\textbf{B}(t)-\textbf{B}(t+\Delta t)$ over a range of time intervals $\Delta t$. For the \textit{MESSENGER} data analysis, fluctuations for 15 successively doubled values of $\Delta t$ ranging from 0.5 to 8192~s were found. The same $\Delta t$ range was used for the \textit{STEREO-B} data analysis with the addition of two lower values (0.125~s and 0.25~s). These timescales span fully the inertial range of the fluctuation spectrum ($10^1$~s~$\lesssim \Delta t \lesssim$~$10^3$~s) and overlap with the low frequency end of the kinetic range ($\Delta t\lesssim10^1$~s).

\begin{figure*}
\epsscale{1.1}
\plotone{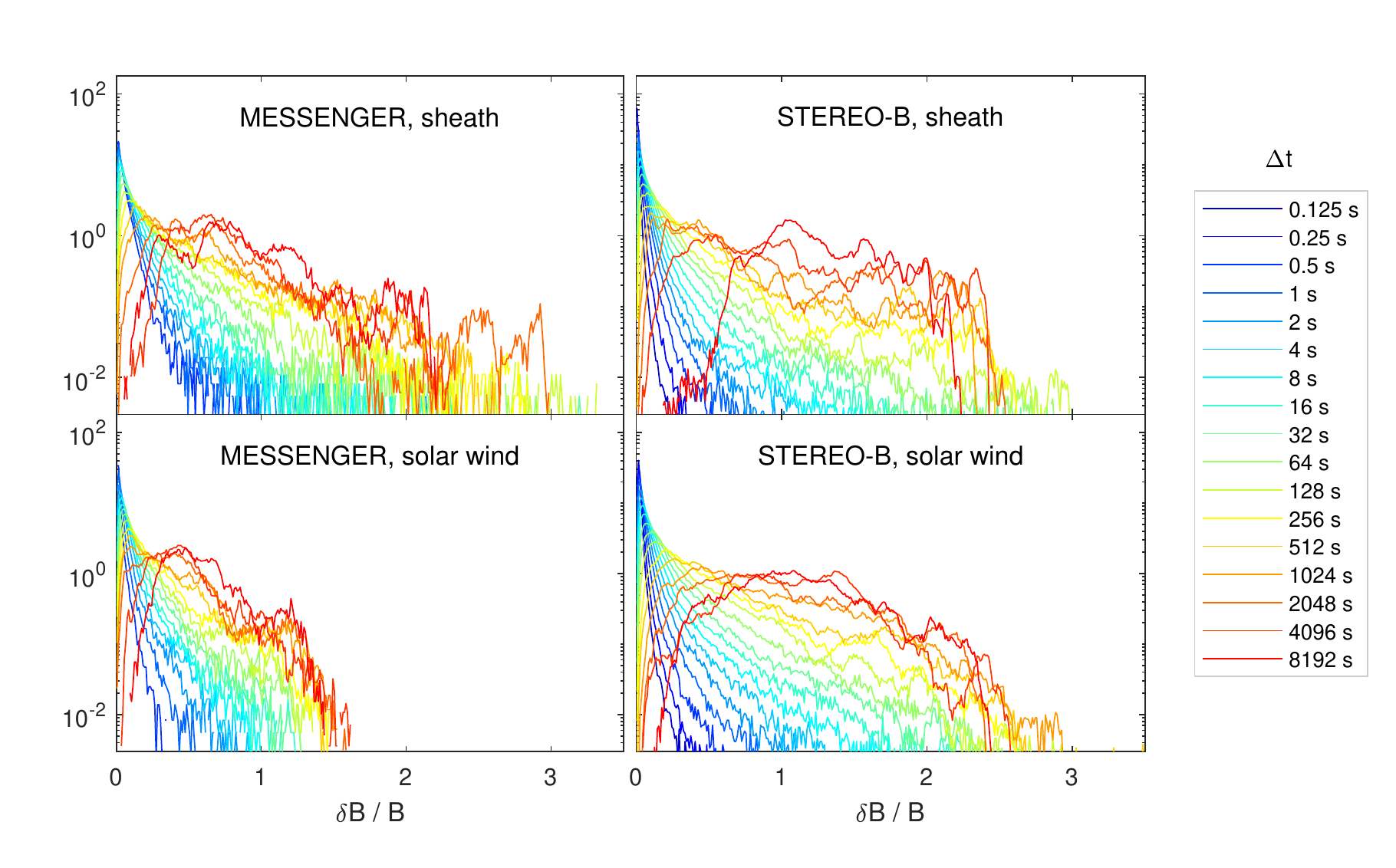}
\caption{Probability distribution functions of $\delta B/B$ across a range of timescales, $\Delta t$. Distributions for the sheath (top panels) and preceding solar wind (bottom panels) at \textit{MESSENGER} (left panels) and \textit{STEREO-B} (right panels) are shown. Note the absence in the \textit{MESSENGER} panels of distributions for the lowest two $\Delta t$ values.}
\label{fig:DelB_dist}
\end{figure*}

Figure~\ref{fig:DelB_dist} shows probability distributions of the normalized fluctuation amplitude, $\delta B/B$, for each $\Delta t$ value. Here the fluctuation amplitude is defined as $\delta B=|\delta\textbf{B}|$, and $B$ is the mean field magnitude between times $t$ and $t+\Delta t$. Distributions in the sheath intervals are shown, as well as distributions in the solar wind immediately ahead of the sheath. The solar wind intervals analyzed were approximately equal in duration to the sheaths at the respective spacecraft. The first 6 minutes of the solar wind interval at \textit{MESSENGER} were excluded because the spacecraft crossed the heliospheric current sheet at this time. Some resampling of the \textit{MESSENGER} solar wind interval was required because higher resolution burst mode measurements were made by the spacecraft within this time period.

Some trends are common to all four distribution sets displayed in Figure~\ref{fig:DelB_dist}. At the smallest timescales, the distributions peak sharply at low $\delta B/B$ values and decay exponentially with increasing $\delta B/B$, consistent with fluctuation amplitudes being small relative to the field magnitude (i.e., $\delta B \ll B$). The distributions are broadly more Gaussian in character and shift towards larger $\delta B/B$ values with increasing timescale. These fluctuation scaling trends are a well established feature of the solar wind plasma \citep{Sorriso-Valvo01} and have recently been demonstrated with $\delta B/B$ distributions in a similar fashion to Figure~\ref{fig:DelB_dist} \citep{Chen15,Matteini18}.

In the sheath at \textit{MESSENGER}, the distributions developed tails extending to $\delta B/B > 2$. These tails were absent in the preceding solar wind. Fluctuations at $\delta B/B > 2$ must be at least partly compressive since a purely Alfv\'enic fluctuation, in which the field magnitude does not change, is limited to $\delta B \leq 2B$. Highly non-Alfv\'enic magnetic holes, in which field magnitudes dip sharply for short intervals, populate the sheath distribution tails at the highest $\delta B/B$ values; a number of magnetic holes are evident in Figure~\ref{fig:B_data} within the sheath time series at both spacecraft. The sheath distributions at \textit{MESSENGER} were also well populated at and just below $\delta B/B = 2$, unlike in the solar wind ahead; fluctuations around $\delta B/B = 2$ are consistent, for example, with the presence of large tangential discontinuities such as current sheets. At \textit{STEREO-B}, there was less difference between the solar wind and sheath distributions, with the distributions being populated at and above $\delta B/B = 2$ in both cases. The apparent evolution in $\delta B/B$ between the solar wind and sheath at \textit{MESSENGER} showed some broad similarities to the evolution seen in the solar wind between \textit{MESSENGER} and \textit{STEREO-B}.

\subsection{Mean Spectral Properties}

\begin{figure}
\epsscale{1.2}
\plotone{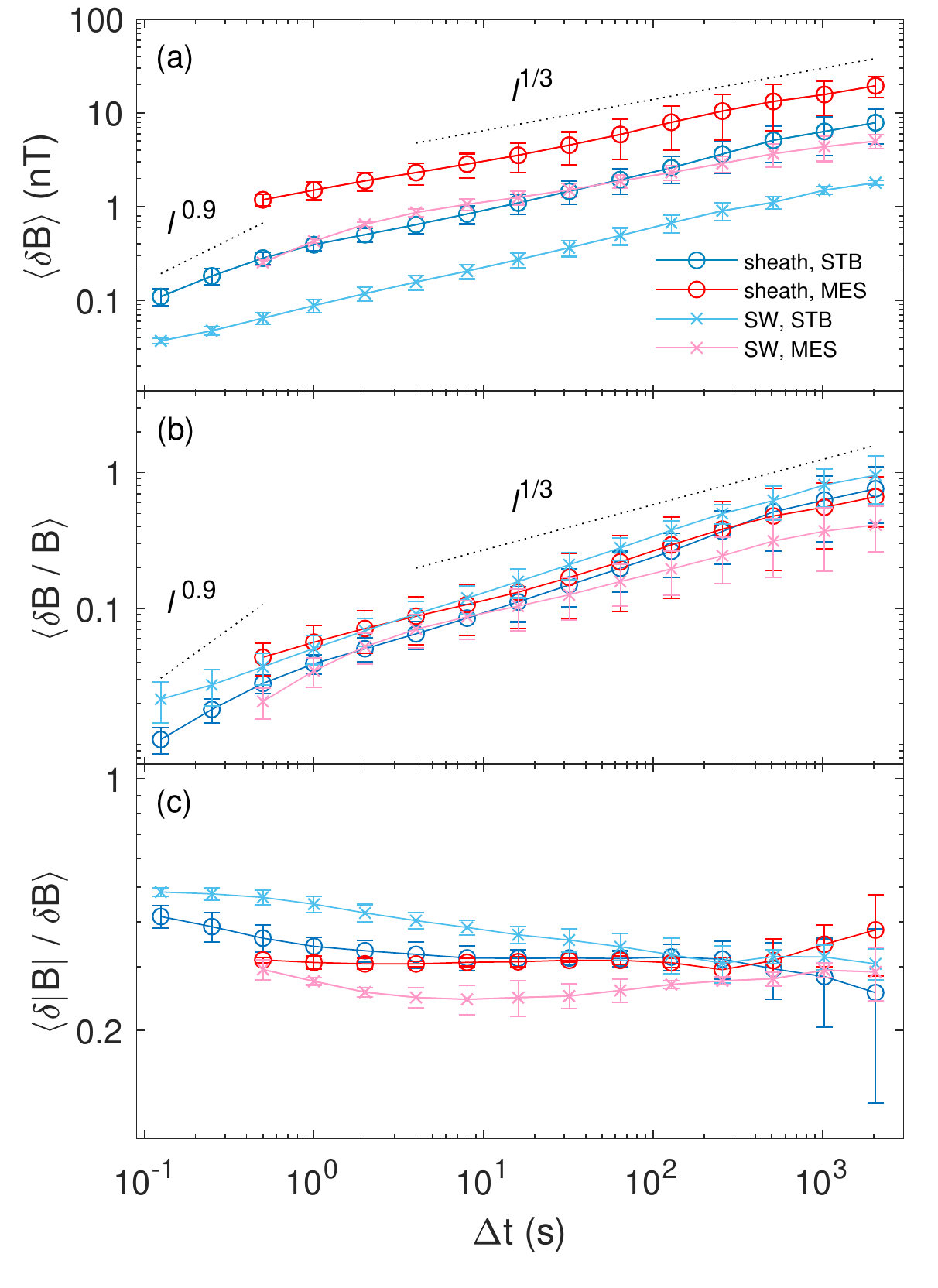}
\caption{(a) Mean fluctuation amplitude, (b) normalized fluctuation amplitude, and (c) fluctuation compressibility, as functions of timescale. Dark blue (pale blue) markers give the values in the sheath (solar wind) at \textit{STEREO-B}, and red (pink) markers give the values in the sheath (solar wind) at \textit{MESSENGER}. The $l^{1/3}$ lines indicate Kolmogorov scaling in the inertial range and the $l^{0.9}$ lines indicate the typically observed kinetic range scaling.}
\label{fig:Mean_values}
\end{figure}

In order to compare further the fluctuations in the sheath and solar wind at the two radial distances, it is useful to consider average parameters across the range of scales. Figure~\ref{fig:Mean_values}a displays mean values of the fluctuation amplitude, $\delta B$, as a function of timescale at both spacecraft. Also displayed are the corresponding values in the solar wind ahead of the sheath. These absolute values are not normalized to the mean field. As expected, fluctuation amplitudes fell with radial distance and were lower in the solar wind than in the sheath. It can be seen that the sheath at 1.08~au and the ambient solar wind at 0.47~au had similar $\langle\delta B\rangle$ values. Error bars in Figure~\ref{fig:Mean_values} are given by the standard deviations of non-overlapping subsampled intervals with duration $\sim 2.5$~hr; this error estimation gives large uncertainties for values at the two largest timescales (4096~s and 8192~s), and so these points have been excluded from the figure.

Figure~\ref{fig:Mean_values}b shows average values of $\delta B/B$ versus timescale. These values correspond to the averages of the distributions shown in Figure~\ref{fig:DelB_dist}. It can be seen that, when normalized to the field magnitude, fluctuation amplitudes collapse to approximately the same line (within error) as a function of timescale. This apparent modulation of field fluctuations by the field magnitude has previously been observed in the solar wind \citep[e.g.,][]{Matteini18}, and we see here that this modulation may also occur in ICME sheath plasma. If the error bars are neglected, it can be seen that $\langle \delta B/B \rangle$ was slightly higher in the sheath at \textit{MESSENGER} than in the preceding solar wind at all scales, while the reverse was true at \textit{STEREO-B}.

The gradients in $\langle\delta B\rangle$ in Figure~\ref{fig:Mean_values}a are related to the nature of the turbulence in the inertial range. For example, in terms of scale \textit{l}, a $\delta{B}\propto~l^{1/3}$ relationship in the inertial range (shown with dashed lines in Figure~\ref{fig:Mean_values}) is equivalent to a \textit{k}-space power spectrum with the familiar Kolmogorov spectral index $\alpha_k = -5/3$; this equivalence may be shown straightforwardly by considering that ${\delta{B}}^2=P(k)\cdot k$, where spectral power $P(k)\propto k^{\alpha_k}$ and $k=1/l$. 

Spectral indices have been determined in the range 4~s~$\leq\Delta t\leq$~2048~s for the various intervals. The sheath fluctuations at 0.47~au had an \textit{l}-space spectral index $\alpha_l = 0.35$, just above the $l^{1/3}$ Kolmogorov scaling, while the upstream solar wind had an index of 0.29. The lower solar wind index may indicate that the turbulence was not fully developed in the interval analyzed, or that the turbulence is better described by a different model. The $\alpha_l = 0.29$ index lies between the indices expected for Kolmogorov and Kraichnan turbulence; the latter, which represents the magnetohydrodynamic extension of the hydrodynamic Kolmogorov theory, predicts a spectral index of $P \propto k^{-3/2} \equiv \delta B \propto l^{1/4}$ in the inertial range. A small part of the increase in $\alpha_l$  between the solar wind and sheath at \textit{MESSENGER} was due to the sheath distribution tails, which were relatively more prominent at larger timescales and hence increased the $\langle \delta B \rangle$ gradient.
 
The solar wind and sheath at 1.08~au had markedly steeper slopes, with indices of $0.40$ and $0.42$, respectively. This steepening relative to 0.47~au may have been due to an enhancement of intermittency, a phenomenon where the fluctuation amplitudes and energy cascade in a turbulent medium is spatially inhomogeneous. We note here the relation of $\langle\delta B\rangle$ to the structure function of the fluctuation amplitude, $\langle|\textbf{B}(t)-\textbf{B}(t+\Delta t)|^{m}\rangle$, where $\langle\delta B\rangle$ is the first-order ($m=1$) function; when Kolmogorov-type turbulence is intermittent, the \textit{l}-space spectral index $\alpha_l$ is expected to be greater than $m/3$ for $m<3$ and less than $m/3$ for $m>3$ \citep[see, e.g., the review by][]{Horbury05}. 

At $\Delta t \lesssim 0.5$~s, there is a steepening of the $\langle \delta B \rangle$ spectrum in the sheath at \textit{STEREO-B}. There is also some steepening in the solar wind spectrum at \textit{MESSENGER}, at $\Delta t \lesssim 2$~s. The steepening is broadly consistent with the $P \propto k^{-2.8}\equiv \delta B \propto l^{0.9}$ power law observed for magnetic fluctuations observed in the kinetic range between ion and electron scales \citep[e.g., see review by][]{Alexandrova13}.

\subsection{Compressibility}

Figure~\ref{fig:Mean_values}c shows mean values of the magnetic compressibility of the fluctuations, $\delta|B|/\delta B$, where $\delta|B|=||\textbf{B}(t)|-|\textbf{B}(t+\Delta t)||$. This quantity gives the mean fraction of the total fluctuation amplitude that involves some compression, i.e., that involves a change in $|B|$. Fast wind, characterized by low compressibility, tends to have $\langle\delta|B|/\delta B\rangle \lesssim 0.2$ at inertial scales, while the more compressible slow wind is observed at higher $\langle\delta|B|/\delta B\rangle$ values \citep{Matteini18}. Compressibility values for the sheath and solar wind displayed in Figure~\ref{fig:Mean_values} are closer to those of the slow wind.

Compressibility reduced in value from the smallest scales ($\Delta t \lesssim 4$~s) to intermediate scales in the sheath at \textit{STEREO-B} and the solar wind intervals at both spacecraft. This trend is in agreement with established findings \citep[e.g.,][]{Chen15}, which have shown solar wind fluctuations to be less compressive in the inertial range than the kinetic. At \textit{MESSENGER}, in contrast, compressibility was approximately flat at small and intermediate scales in the sheath. Compressibility was generally lower in the solar wind than in the sheath at \textit{MESSENGER}, while the reverse was true at \textit{STEREO-B}. There was a convergence in sheath compressibility between \textit{MESSENGER} and \textit{STEREO-B} at intermediate scales (towards $\langle\delta|B|/\delta B\rangle\sim$~0.3) not seen in the solar wind. An overall increase in compressibility with radial distance at small and intermediate scales is evident. Above $\Delta t \gtrsim 512$~s, mean compressibility values were more uncertain.

In making the various sheath--solar wind comparisons, we have thus far not considered any influence of the sheath on the fluctuations in the upstream solar wind. One such influence may have been the SEPs ahead of the shock, observed at \textit{STEREO-B} and by inference also present at \textit{MESSENGER}. SEP protons are known to generate Alfv\'en waves around the proton cyclotron frequency, especially ahead of quasi-parallel shocks \citep[e.g.,][]{Desai12}. We note that there are no significant `humps' in the solar wind spectra around the proton cyclotron frequency or deviations from power-law trends extending into the inertial range in Figure~\ref{fig:Mean_values}a, suggesting that SEP-generated waves for this event were minor relative to the pre-existing turbulent component of the fluctuations at both spacecraft. At \textit{MESSENGER}, where upstream wave activity would have likely been greater due to the shock being quasi-parallel, the interval of solar wind ahead of that previously examined (i.e., the interval between approximately 10 and 5 hr before the shock arrival) displayed marginally lower fluctuation amplitudes and higher compressibility compared to the interval immediately upstream of the shock, consistent with a fall-off in SEP-associated Alfv\'enic fluctuations with distance from the shock. However, there was no difference in the inertial range spectral index between the two solar wind intervals, both having $\alpha_l = 0.29$. Strong Alfv\'enic fluctuations can be seen in the time series for a period of $\sim$11~min immediately ahead of the shock at \textit{MESSENGER}.

\begin{figure*}
\epsscale{1.15}
\plotone{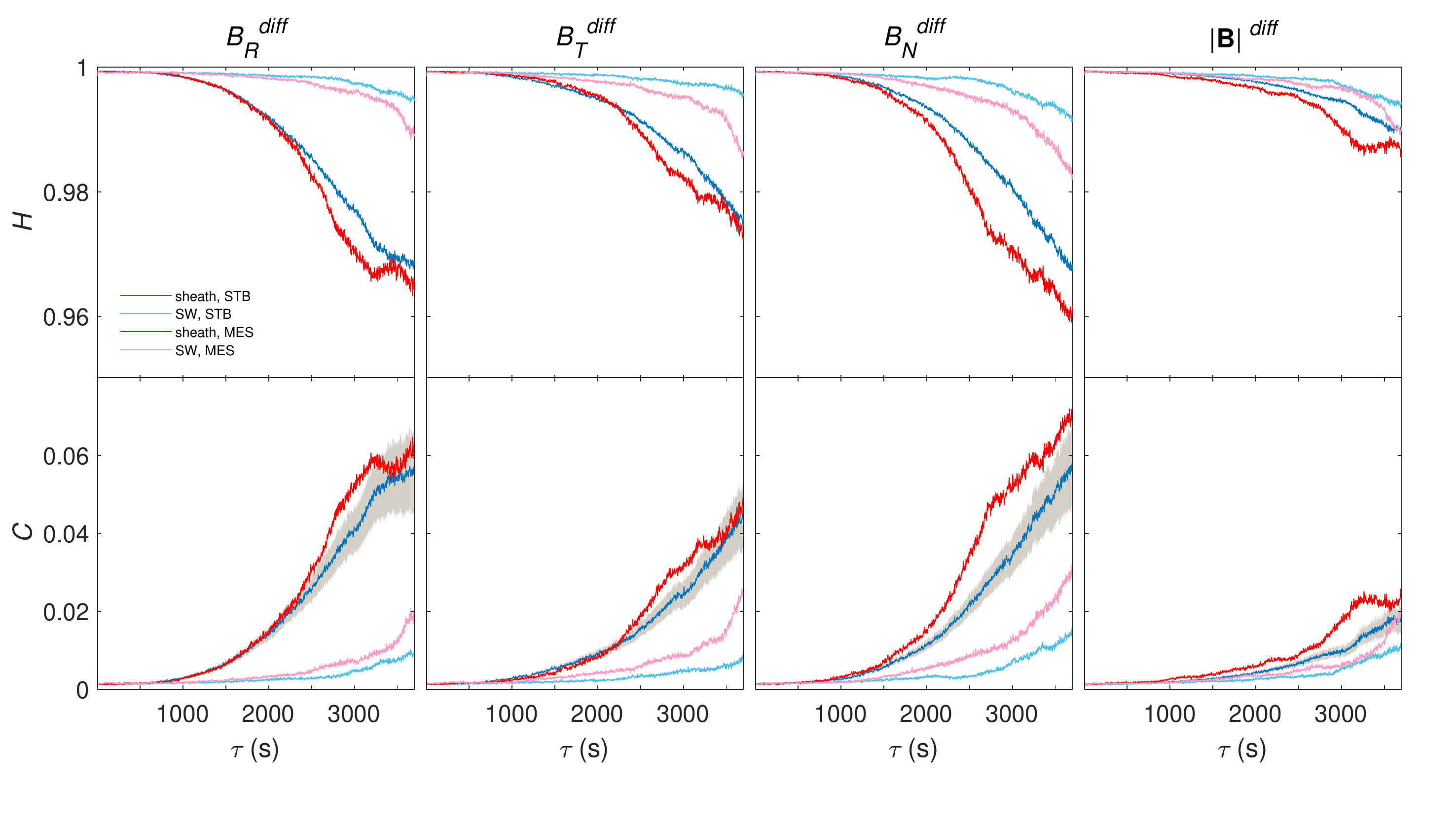}
\caption{Permutation entropy, $H$, and Jensen-Shannon complexity, $C$, in the sheath (solar wind) are shown at \textit{MESSENGER} by the red (pink) lines and at \textit{STEREO-B} by the dark (pale) blue lines. $H$ and $C$ were calculated for the 1~s fluctuations in the RTN coordinates and field magnitude as functions of embedded delay $\tau$. The uncertainty range in $C$ for the sheath at \textit{STEREO-B} is shown by the gray shading.}
\label{fig:CHplots}
\end{figure*}

\section{Entropy and Complexity}

Permutation entropy as defined by \citet{Bandt02} quantifies the probability distribution of permutations (i.e., different amplitude orderings) in a time series. Permutations are computed from subsets of the time series with evenly spaced data points. The number of points in the subset is given by the so-called embedded dimension, $d$, which also defines the number of possible different permutations, $d!$. The spacing of the data points constituting a permutation is expressed by the embedded delay, $\tau$, which effectively defines the time resolution of the subset. Similarly to \citet{Weck15}, \citet{Osmane19}, and \citet{Weygand19}, we have used the normalized Shannon entropy, $H$, to compute the permutation entropy and the Jensen-Shannon complexity, $C$, where $C$ is defined as the product of $H$ and the Jensen divergence \citep{Rosso07}. For exact definitions of the Shannon entropy and the Jensen-Shannon complexity, and further conceptual discussion of permutation entropy and statistical complexity, we direct the reader to the work of \citet{Osmane19} and references therein.

\subsection{Calculation Method}

Entropy and complexity analysis has been applied to the magnetic field time series in the sheath and upstream solar wind at \textit{MESSENGER} and \textit{STEREO-B}. Since the analysis is only strictly valid for stationary time series, i.e., time series in which parameters such as the mean and variance do not themselves change with time, a form of stationarity has been imposed by taking the difference of successive points in the time series; entropy and complexity have been determined for these time series of increments rather than for the time series directly.

Furthermore, in order to make statistically valid comparisons of $H$ and $C$ in the sheath at \textit{MESSENGER} and \textit{STEREO-B} at a particular value of $\tau$, the total number of subsets within the investigated intervals were made equal in size. This was achieved by resampling the data at both spacecraft to a time resolution of 1~s and dividing the sheath interval at \textit{STEREO-B} into four sub-intervals, each with a duration equal to that of the sheath at \textit{MESSENGER}. The sub-intervals were chosen such that they covered 98~\% of the sheath at \textit{STEREO-B} (with the last $\sim10$~min being omitted), and such that two successive sub-intervals had an 80\% overlap. The values of $H$ and $C$ at \textit{STEREO-B} correspond to the means of the values computed for the sub-intervals. This analysis procedure was also applied to the solar wind preceding the sheath. Resampling of the time series to 1~s resolution was applied before obtaining the incremental time series for which $H$ and $C$ were calculated.

\citet{Olivier19} have quantified the effect of data gaps on the probability vector used to compute $H$ and $C$. Following the suggestion of these authors, we have omitted subsets with missing data points ($\sim$0.1\% of total subsets). Olivier et al. also found that data re-sampling has some affect on the values of $H$ and $C$ at low $\tau$, and recommend using values of $\tau$ at least twenty times larger than the re-sampling frequency, i.e., $\tau=20$~s for the 1~s sampling we have used. However, we have also calculated $H$ and $C$ at $\tau$ values below this limit for completeness.

An embedded dimension of $d=5$ was chosen for the analysis, and embedded delay $\tau$ was varied between $\tau=2$~s and $\tau=3700$~s in steps of 1~s. The total time durations of the subsets giving the permutations thus ranged from 4~s to 4~hr 7~min. The number of subsets is equal to $N-(d-1)\tau$, where $N$ is the number of data points within the data interval analyzed; this gave a range in the total number of subsets of $3550<N-(d-1)\tau<18336$. To further verify the statistical robustness of the analysis, we confirmed that the studied data intervals satisfied the criteria $N/d!>10$ and $\sqrt{d!/(N-(d-1)\tau)}<0.2$ \citep{Osmane19} for every value of $\tau$.

\subsection{Results}

Figure~\ref{fig:CHplots} shows the values of $H$ and $C$ versus $\tau$ for the 1~s fluctuation amplitudes of the magnetic field magnitude and RTN components. The sheath interval and preceding solar wind at each spacecraft have been analyzed separately as in previous sections. The uncertainty in $C$ defined by \citet{Weygand19}, $\sqrt{d!/[N-(d-1)\tau]}$, is shown in Figure~\ref{fig:CHplots} by the gray shading for the \textit{STEREO-B} sheath values; this uncertainty rises from 8\% at the lowest $\tau$ value to 18\% at the highest value, with similar uncertainties (not shown in the figure) for all other $C$ lines.

It can be seen that $H$ was high ($\gtrsim 0.96$) and $C$ was low ($\lesssim 0.07$) for all $\tau$ in all cases, consistent with the presence of highly stochastic fluctuations. At $\tau \lesssim 800$~s, $H$ was near unity and $C$ near zero. With $\tau$ increasing from $\sim 800$~s to $\sim 2500-2700$~s, there were power-law falls in $H$ and rises in $C$ in the sheath field components, with less monotonic continuations of these trends at $\tau\gtrsim 2700$~s. In the upstream solar wind, $H$ also fell and $C$ also rose in the field components with increasing $\tau$ but the trends were much less pronounced, with $H$ remaining at higher values and $C$ remaining at lower values than in the sheath. Roughly similar trends were also seen for the field magnitude in the sheath and solar wind, but the fall in $H$ and rise in $C$ with $\tau$ was much weaker for the sheath magnitude than for the sheath components.

There is also some evidence of a radial evolution in entropy and complexity in Figure~\ref{fig:CHplots}. At larger values of $\tau$, $H$ was generally higher in the sheath at 1.08~au than in the sheath at 0.47~au. Likewise, $C$ was lower in the sheath at 1.08~au than at 0.47~au at larger $\tau$. Similar radial trends were seen for the upstream solar wind observed at the two radial distances. At lower values of $\tau$ ($\lesssim 2000$~s), there was generally little or no dependence of $H$ and $C$ on radial distance.

\citet{Weygand19} consider the impact of instrument noise when measuring radial variations in complexity. Fluxgate magnetometers such as those onboard \textit{MESSENGER} and \textit{STEREO-B} have a pink (i.e., $1/f$) noise floor. This instrument noise is highly stochastic and has low complexity. As fluctuation amplitudes in the solar wind fall with radial distance, signal-to-noise ratios will also tend to fall (assuming the spacecraft sampling fluctuations at the different distances have similar noise levels), and so complexity may reduce as the stochastic pink noise becomes a relatively more significant part of the measured field. A similar effect may be seen in single-spacecraft measurements upstream and downstream of a shock. However, since both spacecraft had very low noise floors well below inner-heliospheric fluctuation amplitudes, with intrinsic 1-Hz noise levels of $<20$~pT~$\sqrt{}$~Hz at \textit{MESSENGER} \citep{Anderson07} and  $<10$~pT~$\sqrt{}$~Hz at \textit{STEREO-B} \citep{Acuna08}, variations in complexity values are likely dominated by variations in the signal properties.

\section{Planar Magnetic Structure}

\begin{figure*}
\epsscale{1.18}
\plotone{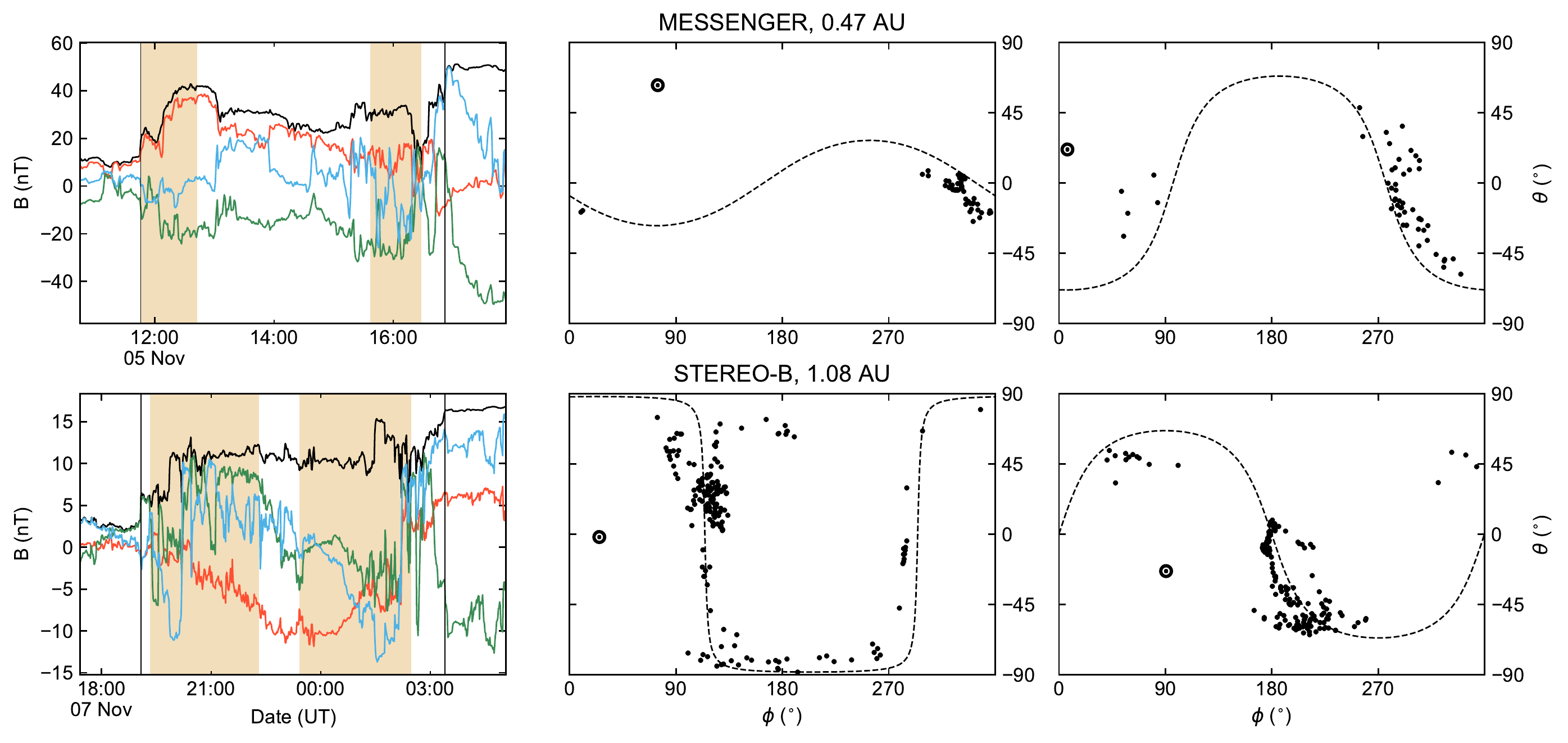}
\caption{Planar structuring in the sheath at \textit{MESSENGER} (top panels) and \textit{STEREO-B} (bottom panels). Shading overlaying the 1-minute magnetic field data in the left-hand panels shows the planar structuring intervals. Center and right-hand panels show $\theta$-$\phi$ diagrams for the planar intervals near the shock and flux rope, respectively; the dashed lines in these panels define the planes and the $\odot$ symbols indicate the anti-sunward plane normals.}
\label{fig:pms_figure}
\end{figure*}

We now consider in detail one feature of the sheath's large-scale spatial structure, namely, magnetic planarity. The magnetic field in ICME sheaths may form planar sheets in which successive magnetic field vectors vary in direction within the plane but not normal to it \citep{Nakagawa89}.

Planar structuring in the sheath has been identified using the methods of \citet{Palmerio16}. In outline, the 1~au search technique involves the cumulative removal of successive 5-minute increments of data from the sheath until planar structure is identified in the remaining data. Minimum variance analysis is used to identify any planar field variation, with a minimum duration threshold for positive identification set at 1 hr. To allow for typically shorter sheath durations at 0.5~au, the removal increment and minimum duration threshold were scaled down to 3~minutes and 37~minutes, respectively. Data at both spacecraft were resampled to a resolution of 1~minute for this analysis.

Two intervals of planar structuring were identified at each radial distance. These intervals are shown by the shaded regions in the left-hand panels of Figure~\ref{fig:pms_figure}. At both spacecraft, the intervals were located immediately behind the leading shock (PMS1) and ahead of the flux rope (PMS2), with respective durations of 57 and 51~minutes at \textit{MESSENGER}, and durations of 179 and 184~minutes at \textit{STEREO-B}. The center and right-hand panels show $\theta$-$\phi$ diagrams for PMS1 and PMS2, respectively; these diagrams, which plot the latitude angles ($\theta$) versus the longitude angles ($\phi$) of the magnetic field vectors in RTN coordinates, show clustering of points around lines that define planes. The planar intervals grew in length (both in absolute duration and relative to the sheath duration) with radial distance and became more clearly defined, as illustrated by the greater spread of points along the plane lines in the $\theta$-$\phi$ diagrams. Development of ICME sheath planar structuring with radial distance in the outer heliosphere has previously been reported by \citet{Intriligator08}. A recent study by \citet{Lugaz20} identified a planar sheath structure observed at two aligned spacecraft that expanded with radial distance at the same rate as the driving ICME.

It is tempting to ascribe PMS1 to shock-aligned compression \citep{Jones02} and PMS2 to field line draping around the flux rope \citep{Farrugia90}. The angle between the PMS1 normal and shock normal is relatively small at both \textit{MESSENGER} ($33^{\circ}$) and \textit{STEREO-B} ($12^{\circ}$), consistent with shock alignment. We have used the IP Shocks and coplanarity theorem shock orientations at \textit{STEREO-B} and \textit{MESSENGER}, respectively, to make this comparison. If PMS2 is due to flux rope draping, the plane normal and flux rope axis orientation would likely be separated by large acute angles and this is indeed the case, with separations of $\sim 73^{\circ}$ at \textit{MESSENGER} and $\sim 57^{\circ}$ at \textit{STEREO-B}. The flux rope orientations used to make this comparison were determined by \citet{Good19}.

\section{Discussion and Conclusion} 

A range of processes contribute to the radial evolution of magnetic field fluctuations in ICME sheaths. As sheaths propagate, they tend to grow in radial width \citep{Janvier19,Salman20} as upstream plasma is swept downstream of the shock and added to the sheath material. Fluctuations found in sheaths thus comprise pre-existing fluctuations from the solar wind that become compressed, and also new fluctuations generated locally within the sheath. The upstream solar wind may be spatially inhomogeneous and its properties may evolve with distance, meaning that the nature of the fluctuations being added to the sheath will also change with distance. Furthermore, an ICME's passage can itself modify upstream wind fluctuations; fluctuations can be produced, for example, by the acceleration of SEPs ahead of ICME shocks. SEP-generated waves will generally be more significant closer to the Sun, since quasi-parallel shocks (more prevalent at sub-1~au distances) accelerate SEPs much more effectively than quasi-perpendicular shocks. Properties of the fluctuations injected downstream by shocks also change systematically with shock geometry.

In this work, we have performed the first analysis of magnetic field fluctuations $\delta B$ in an ICME sheath at two aligned observation points in the inner heliosphere. Probability distributions of $\delta B/B$ have been determined as functions of timescale (Figure~\ref{fig:DelB_dist}), in both the sheath and preceding solar wind. Distributions were determined for fluctuation timescales spanning the inertial range and large-scale end of the kinetic range.

Significant differences were seen in the distributions between the two radial distances. At 0.47~au, the sheath distributions displayed tails at high $\delta B/B$ values, consistent with the development of sharp field discontinuities and large-angle field rotations, magnetic holes, and other highly compressive structures. These tails were absent in the upstream solar wind. At 1.08~au, in contrast, there was much greater similarity in the form of the sheath and solar wind distributions, with the distributions extending beyond $\delta B/B = 2$ in both cases. The shift in the $\delta B/B$ distributions towards higher $\delta B/B$ values seen with the solar wind-sheath transition at 0.47~au was qualitatively similar to the `aging effect' shift seen in the upstream solar wind between the two radial distances.

Mean values of the distributions (Figure~\ref{fig:Mean_values}b) are consistent with the interpretation above, i.e., a shift towards higher $\langle \delta B/B \rangle$ values at all timescales in the sheath at 0.47~au compared to the upstream wind, and a similar shift in the upstream wind with heliocentric distance. A slight drop in $\langle \delta B/B \rangle$ from the solar wind to sheath at 1.08~au can also be seen. However, changes in $\langle \delta B/B \rangle$ between the various intervals are generally small and error bars, determined from interval sub-sampling, are relatively large. We note here the value in considering $\delta B/B$ distributions rather than their mean values alone: distributions allow the dominant core population of fluctuations at $\delta B/B < 2$, mostly comprising incompressible Alfv\'enic turbulence, to be be distinguished from the highly compressive fluctuations that populate the tails. This distinction is lost with averaging.

Although $\langle \delta B/B \rangle$ only differed by small or statistically negligible amounts between solar wind and sheath or with radial distance, the mean compressibility (Figure~\ref{fig:Mean_values}c) varied more significantly. Thus, there were changes in the composition of the fluctuations (in terms of relative Alfv\'enicity and compressibility) without correspondingly significant changes in the total normalized amplitudes, $\langle\delta B/B\rangle$, which give the sum of Alfv\'enic and compressive components. At kinetic and inertial--range scales, sheath compressibility was higher at 0.47~au and lower at 1.08~au relative to the preceding solar wind in each case. There was also a general trend towards greater compressibility with radial distance in both the sheath and solar wind. The reduction in compressibilty from solar wind to sheath at 1.08~au was consistent with an enhancement in Alfv\'enic relative to compressive fluctuations in the sheath. Overall sheath compressibility may have been modified be the leading shock; the shock was quasi-parallel at 0.47~au, and quasi-parallel shocks are known to inject compressive fluctuations into the downstream plasma and enhance compressive fluctuation power. \citet{Moissard19}, for example, found that sheaths preceded by quasi-parallel shocks tended to have lower power anisotropies and a more equal distribution of power between Alfv\'enic and compressive fluctuations than sheaths preceded by quasi-perpendicular shocks at 1~au, although this dependence was not strong and based on a relatively small number of quasi-parallel events. Furthermore, large-amplitude compressive fluctuations were clustered near the sheath trailing edge at both spacecraft, suggesting that processes associated with the sheath-ICME boundary also played a role in modifying overall compressibility.

Power law scalings of fluctuations at inertial range timescales in the sheath and solar wind at both spacecraft (Figure~\ref{fig:Mean_values}a) were consistent with the presence of turbulence. The spectral slope in the solar wind at 0.47~au was relatively shallow, suggestive of either an under-developed cascade or the presence of Kraichnan-like rather than Kolmogorov-like turbulence. Steepening between the solar wind and sheath at 0.47~au may have have been due to cascade development or growth in intermittency; steepening between the spacecraft, to slopes at 1.08~au significantly steeper than that of Kolomogorov turbulence, may likewise have been due to increased intermittency. Greater intermittency in magnetic fields and the correspondingly steeper spectral slopes have, for example, been associated with an increased presence of current sheets in the plasma \citep{Li11}. In the distributions shown in Figure~\ref{fig:DelB_dist}, strong current sheets would appear as incompressible enhancements approaching $\delta B/B = 2$ (i.e., $180^{\circ}$ flips in the field direction); such enhancements were present in both sheath intervals and the solar wind interval at 1.08~au. The spectral steepening may also partly have been due to sampling variation: \citet{Borovsky12}, for example, found a normally-distributed variation in $\alpha_k$ centered on the Kolmogorov index when sampling relatively short data intervals such as those used in this study.

The first analysis of entropy and Jensen-Shannon complexity in an ICME sheath observed at different heliocentric distances has been presented in this work (Figure~\ref{fig:CHplots}); this also represents, as far as the authors are aware, the first such analysis upstream and downstream of a shock. At larger scales, a trend towards reducing complexity in the patterns of 1~s fluctuation amplitudes with radial distance was found, along with an even more pronounced trend of increased complexity in the sheath intervals relative to the upstream solar wind. This latter finding is consistent with and analogous to the generally increased complexity in stream interaction regions relative to the solar wind that was found by \citet{Weygand19}. A greater large-scale complexity in sheaths compared to the solar wind is in agreement with our understanding of sheath plasma, i.e., a plasma that contains a variable mix of coherent, ordered structures and random, disordered fluctuations. As sheath structures become less prevalent with reducing scale, complexity decreases towards values found in the relatively unstructured solar wind. Also, the lower complexity values of the compressive field magnitude fluctuations suggest that they are comparatively less structured and more stochastic than the Alfv\'enic sheath fluctuations, which contribute to the field component complexity.

Two intervals of planar magnetic structure were identified in the sheath (Figure~\ref{fig:pms_figure}), both of which grew significantly as a fraction of the total sheath interval with radial distance. The fractional growth in the near-shock planar structure may simply have been due to the steady accumulation and compression of solar wind being added to the sheath with distance, while the near-rope structure may have grown with accumulated draping to the sheath rear. Growth in the near-shock structure may have been aided by the transition of the shock from a quasi-parallel to quasi-perpendicular geometry, since planar structures form more easily behind quasi-perpendicular shocks \citep{Jones02,Palmerio16}. The increase in duration of both intervals with radial distance may also partly have been due to expansion \citep{Lugaz20}.

In the preceding discussion, comparisons have been made between upstream solar wind intervals and downstream sheath intervals, and differences between the intervals have been used to infer upstream-to-downstream evolution in fluctuation properties. Although some differences were almost certainly evolutionary (e.g., increased downstream amplitudes and spectral slope variations), other differences may have been due to spatial inhomogeneities. We emphasize finally that the trends found in this case study may not be indicative of general trends, and further spacecraft line-up studies are required to build a statistical picture. New opportunities to perform studies of ICME sheath evolution with heliocentric distance will hopefully be provided by \textit{Parker Solar Probe} and \textit{Solar Orbiter}. ICME sheaths may also be probed by these spacecraft at early and previously unobserved stages of development close to the Sun.


\acknowledgments

We thank the \textit{MESSENGER} and \textit{STEREO} instrument teams for the data used in this study. Data were obtained from the CDAWeb archive \linebreak (https://cdaweb.sci.gsfc.nasa.gov). S.G. and E.K. are supported by Academy of Finland grant 310445 (SMASH), and M.A.-L. and E.K. are supported by funding from the European Research Council (ERC) under the European Union’s Horizon 2020 research and innovation programme grant 724391 (SolMAG). All authors are additionally supported by Academy of Finland Centre of Excellence grant 312390 (FORESAIL). E.P. acknowledges the NASA Living With a Star Jack Eddy Postdoctoral Fellowship Program, administered by UCAR's Cooperative Programs for the Advancement of Earth System Science (CPAESS) under award no. NNX16AK22G. We also wish to thank the anonymous reviewer, whose thoughtful and constructive comments have lead to a much improved manuscript.






\bibliography{bibliography.bib} 

\begin{thebibliography}{}
\expandafter\ifx\csname natexlab\endcsname\relax\def\natexlab#1{#1}\fi
\providecommand{\url}[1]{\href{#1}{#1}}
\providecommand{\dodoi}[1]{doi:~\href{http://doi.org/#1}{\nolinkurl{#1}}}
\providecommand{\doeprint}[1]{\href{http://ascl.net/#1}{\nolinkurl{http://ascl.net/#1}}}
\providecommand{\doarXiv}[1]{\href{https://arxiv.org/abs/#1}{\nolinkurl{https://arxiv.org/abs/#1}}}

\bibitem[{{Abraham-Shrauner} \& {Yun}(1976)}]{Abraham-Shrauner76}
{Abraham-Shrauner}, B., \& {Yun}, S.~H. 1976, \jgr, 81, 2097,
  \dodoi{10.1029/JA081i013p02097}

\bibitem[{{Acu{\~n}a} {et~al.}(2008){Acu{\~n}a}, {Curtis}, {Scheifele},
  {Russell}, {Schroeder}, {Szabo}, \& {Luhmann}}]{Acuna08}
{Acu{\~n}a}, M.~H., {Curtis}, D., {Scheifele}, J.~L., {et~al.} 2008, \ssr, 136,
  203, \dodoi{10.1007/s11214-007-9259-2}

\bibitem[{Ala-Lahti {et~al.}(2019)Ala-Lahti, Kilpua, Sou{\v c}ek, Pulkkinen, \&
  Dimmock}]{Ala-Lahti19}
Ala-Lahti, M., Kilpua, E. K.~J., Sou{\v c}ek, J., Pulkkinen, T.~I., \& Dimmock,
  A.~P. 2019, \jgr, \dodoi{10.1029/2019JA026579}

\bibitem[{{Ala-Lahti} {et~al.}(2018){Ala-Lahti}, {Kilpua}, {Dimmock}, {Osmane},
  {Pulkkinen}, \& Sou{\v c}ek}]{Ala-Lahti18}
{Ala-Lahti}, M.~M., {Kilpua}, E. K.~J., {Dimmock}, A.~P., {et~al.} 2018,
  \angeo, 36, 793, \dodoi{10.5194/angeo-36-793-2018}

\bibitem[{{Alexandrova} {et~al.}(2013){Alexandrova}, {Chen}, {Sorriso-Valvo},
  {Horbury}, \& {Bale}}]{Alexandrova13}
{Alexandrova}, O., {Chen}, C.~H.~K., {Sorriso-Valvo}, L., {Horbury}, T.~S., \&
  {Bale}, S.~D. 2013, \ssr, 178, 101, \dodoi{10.1007/s11214-013-0004-8}

\bibitem[{{Anderson} {et~al.}(2007){Anderson}, {Acu{\~n}a}, {Lohr},
  {Scheifele}, {Raval}, {Korth}, \& {Slavin}}]{Anderson07}
{Anderson}, B.~J., {Acu{\~n}a}, M.~H., {Lohr}, D.~A., {et~al.} 2007, \ssr, 131,
  417, \dodoi{10.1007/s11214-007-9246-7}

\bibitem[{Bandt \& Pompe(2002)}]{Bandt02}
Bandt, C., \& Pompe, B. 2002, \prl, 88, 174102,
  \dodoi{10.1103/PhysRevLett.88.174102}

\bibitem[{{Bavassano} {et~al.}(1982){Bavassano}, {Dobrowolny}, {Mariani}, \&
  {Ness}}]{Bavassano82}
{Bavassano}, B., {Dobrowolny}, M., {Mariani}, F., \& {Ness}, N.~F. 1982, \jgr,
  87, 3617, \dodoi{10.1029/JA087iA05p03617}

\bibitem[{{Borovsky}(2012)}]{Borovsky12}
{Borovsky}, J.~E. 2012, \jgr, 117, A05104, \dodoi{10.1029/2011JA017499}

\bibitem[{{Bruno} \& {Trenchi}(2014)}]{Bruno14}
{Bruno}, R., \& {Trenchi}, L. 2014, \apjl, 787, L24,
  \dodoi{10.1088/2041-8205/787/2/L24}

\bibitem[{{Chen} {et~al.}(2015){Chen}, {Matteini}, {Burgess}, \&
  {Horbury}}]{Chen15}
{Chen}, C.~H.~K., {Matteini}, L., {Burgess}, D., \& {Horbury}, T.~S. 2015,
  \mnras, 453, L64, \dodoi{10.1093/mnrasl/slv107}

\bibitem[{{Colburn} \& {Sonett}(1966)}]{Colburn66}
{Colburn}, D.~S., \& {Sonett}, C.~P. 1966, \ssr, 5, 439,
  \dodoi{10.1007/BF00240575}

\bibitem[{{D'Amicis} {et~al.}(2010){D'Amicis}, {Bruno}, {Pallocchia},
  {Bavassano}, {Telloni}, {Carbone}, \& {Balogh}}]{D'Amicis10}
{D'Amicis}, R., {Bruno}, R., {Pallocchia}, G., {et~al.} 2010, \apj, 717, 474,
  \dodoi{10.1088/0004-637X/717/1/474}

\bibitem[{{Desai} {et~al.}(2012){Desai}, {Dayeh}, {Smith}, {Mason}, \&
  {Lee}}]{Desai12}
{Desai}, M., {Dayeh}, M., {Smith}, C., {Mason}, G., \& {Lee}, M. 2012, in
  American Institute of Physics Conference Series, Vol. 1436, American
  Institute of Physics Conference Series, ed. J.~{Heerikhuisen}, G.~{Li},
  N.~{Pogorelov}, \& G.~{Zank}, 110--115

\bibitem[{{Farrugia} {et~al.}(1990){Farrugia}, {Dunlop}, {Geurts}, {Balogh},
  {Southwood}, {Bryant}, {Neugebauer}, \& {Etemadi}}]{Farrugia90}
{Farrugia}, C.~J., {Dunlop}, M.~W., {Geurts}, F., {et~al.} 1990, \grl, 17,
  1025, \dodoi{10.1029/GL017i008p01025}

\bibitem[{{Good} {et~al.}(2019){Good}, {Kilpua}, {LaMoury}, {Forsyth},
  {Eastwood}, \& {M{\"o}stl}}]{Good19}
{Good}, S.~W., {Kilpua}, E.~K.~J., {LaMoury}, A.~T., {et~al.} 2019, \jgr, 124,
  \dodoi{10.1029/2019JA026475}

\bibitem[{{Horbury} {et~al.}(2005){Horbury}, {Forman}, \&
  {Oughton}}]{Horbury05}
{Horbury}, T.~S., {Forman}, M.~A., \& {Oughton}, S. 2005, \ppcf, 47, B703,
  \dodoi{10.1088/0741-3335/47/12B/S52}

\bibitem[{{Howes} {et~al.}(2014){Howes}, {Klein}, \& {TenBarge}}]{Howes14}
{Howes}, G.~G., {Klein}, K.~G., \& {TenBarge}, J.~M. 2014, \apj, 789, 106,
  \dodoi{10.1088/0004-637X/789/2/106}

\bibitem[{{Intriligator} {et~al.}(2008){Intriligator}, {Rees}, \&
  {Horbury}}]{Intriligator08}
{Intriligator}, D.~S., {Rees}, A., \& {Horbury}, T.~S. 2008, \jgr, 113, A05102,
  \dodoi{10.1029/2007JA012699}

\bibitem[{{Janvier} {et~al.}(2019){Janvier}, {Winslow}, {Good}, {Bonhomme},
  {D{\'e}moulin}, {Dasso}, {M{\"o}stl}, {Lugaz}, {Amerstorfer}, {Soubri{\'e}},
  \& {Boakes}}]{Janvier19}
{Janvier}, M., {Winslow}, R.~M., {Good}, S., {et~al.} 2019, \jgr, 124, 812,
  \dodoi{10.1029/2018JA025949}

\bibitem[{{Jones} {et~al.}(2002){Jones}, {Rees}, {Balogh}, \&
  {Forsyth}}]{Jones02}
{Jones}, G.~H., {Rees}, A., {Balogh}, A., \& {Forsyth}, R.~J. 2002, \grl, 29,
  1520, \dodoi{10.1029/2001GL014110}

\bibitem[{{Kilpua} {et~al.}(2017){Kilpua}, {Koskinen}, \&
  {Pulkkinen}}]{Kilpua17}
{Kilpua}, E., {Koskinen}, H. E.~J., \& {Pulkkinen}, T.~I. 2017, \lrsp, 14, 5,
  \dodoi{10.1007/s41116-017-0009-6}

\bibitem[{{Li} {et~al.}(2011){Li}, {Miao}, {Hu}, \& {Qin}}]{Li11}
{Li}, G., {Miao}, B., {Hu}, Q., \& {Qin}, G. 2011, \prl, 106, 125001,
  \dodoi{10.1103/PhysRevLett.106.125001}

\bibitem[{{Li} {et~al.}(2019){Li}, {Yang}, {Wu}, \& {Wang}}]{Li19}
{Li}, Q.~H., {Yang}, L., {Wu}, D.~J., \& {Wang}, T.~Y. 2019, \apj, 874, 55,
  \dodoi{10.3847/1538-4357/ab06f7}

\bibitem[{{Liu} {et~al.}(2006){Liu}, {Richardson}, {Belcher}, {Kasper}, \&
  {Skoug}}]{Liu06}
{Liu}, Y., {Richardson}, J.~D., {Belcher}, J.~W., {Kasper}, J.~C., \& {Skoug},
  R.~M. 2006, \jgr, 111, A09108, \dodoi{10.1029/2006JA011723}

\bibitem[{{Lugaz} {et~al.}(2020){Lugaz}, {Winslow}, \& {Farrugia}}]{Lugaz20}
{Lugaz}, N., {Winslow}, R.~M., \& {Farrugia}, C.~J. 2020, \jgr, 125,
  e2019JA027213, \dodoi{10.1029/2019JA027213}

\bibitem[{{Marsch} \& {Tu}(1990)}]{Marsch90}
{Marsch}, E., \& {Tu}, C.~Y. 1990, \jgr, 95, 8211,
  \dodoi{10.1029/JA095iA06p08211}

\bibitem[{{Matteini} {et~al.}(2018){Matteini}, {Stansby}, {Horbury}, \&
  {Chen}}]{Matteini18}
{Matteini}, L., {Stansby}, D., {Horbury}, T.~S., \& {Chen}, C.~H.~K. 2018,
  \apj, 869, L32, \dodoi{10.3847/2041-8213/aaf573}

\bibitem[{{Moissard} {et~al.}(2019){Moissard}, {Fontaine}, \&
  {Savoini}}]{Moissard19}
{Moissard}, C., {Fontaine}, D., \& {Savoini}, P. 2019, \jgr, 124, 8208,
  \dodoi{10.1029/2019JA026952}

\bibitem[{{Mu{\~n}oz} {et~al.}(2018){Mu{\~n}oz}, {Dom{\'\i}nguez}, {Alejand ro
  Valdivia}, {Good}, {Nigro}, \& {Carbone}}]{Munoz18}
{Mu{\~n}oz}, V., {Dom{\'\i}nguez}, M., {Alejand ro Valdivia}, J., {et~al.}
  2018, \npgeo, 25, 207, \dodoi{10.5194/npg-25-207-2018}

\bibitem[{{Nakagawa} {et~al.}(1989){Nakagawa}, {Nishida}, \&
  {Saito}}]{Nakagawa89}
{Nakagawa}, T., {Nishida}, A., \& {Saito}, T. 1989, \jgr, 94, 11761,
  \dodoi{10.1029/JA094iA09p11761}

\bibitem[{{Olivier} {et~al.}(2019){Olivier}, {Engelbrecht}, \&
  {Strauss}}]{Olivier19}
{Olivier}, C.~P., {Engelbrecht}, N.~E., \& {Strauss}, R.~D. 2019, \jgr, 124, 4,
  \dodoi{10.1029/2018JA026102}

\bibitem[{{Osmane} {et~al.}(2019){Osmane}, {Dimmock}, \&
  {Pulkkinen}}]{Osmane19}
{Osmane}, A., {Dimmock}, A.~P., \& {Pulkkinen}, T.~I. 2019, \jgr, 124, 2541,
  \dodoi{10.1029/2018JA026248}

\bibitem[{{Palmerio} {et~al.}(2016){Palmerio}, {Kilpua}, \&
  {Savani}}]{Palmerio16}
{Palmerio}, E., {Kilpua}, E. K.~J., \& {Savani}, N.~P. 2016, \angeo, 34, 313,
  \dodoi{10.5194/angeo-34-313-2016}

\bibitem[{{Riazantseva} {et~al.}(2019){Riazantseva}, {Rakhmanova}, {Zastenker},
  {Yermolaev}, \& {Lodkina}}]{Riazantseva19}
{Riazantseva}, M.~O., {Rakhmanova}, L.~S., {Zastenker}, G.~N., {Yermolaev},
  Y.~I., \& {Lodkina}, I.~G. 2019, \geae, 59, 127,
  \dodoi{10.1134/S0016793219020117}

\bibitem[{Rosso {et~al.}(2007)Rosso, Larrondo, Martin, Plastino, \&
  Fuentes}]{Rosso07}
Rosso, O.~A., Larrondo, H.~A., Martin, M.~T., Plastino, A., \& Fuentes, M.~A.
  2007, \prl, 99, 154102, \dodoi{10.1103/PhysRevLett.99.154102}

\bibitem[{{Salman} {et~al.}(2020){Salman}, {Winslow}, \& {Lugaz}}]{Salman20}
{Salman}, T.~M., {Winslow}, R.~M., \& {Lugaz}, N. 2020, \jgr, 125,
  e2019JA027084, \dodoi{10.1029/2019JA027084}

\bibitem[{{Schwartz}(1998)}]{Schwartz98}
{Schwartz}, S.~J. 1998, ISSI Scientific Reports Series, 1, 249

\bibitem[{{Schwartz} \& {Marsch}(1983)}]{Schwartz83}
{Schwartz}, S.~J., \& {Marsch}, E. 1983, \jgr, 88, 9919,
  \dodoi{10.1029/JA088iA12p09919}

\bibitem[{{Siscoe} \& {Odstrcil}(2008)}]{Siscoe08}
{Siscoe}, G., \& {Odstrcil}, D. 2008, \jgr, 113, A00B07,
  \dodoi{10.1029/2008JA013142}

\bibitem[{{Sorriso-Valvo} {et~al.}(2001){Sorriso-Valvo}, {Carbone}, {Giuliani},
  {Veltri}, {Bruno}, {Antoni}, \& {Martines}}]{Sorriso-Valvo01}
{Sorriso-Valvo}, L., {Carbone}, V., {Giuliani}, P., {et~al.} 2001, \planss, 49,
  1193, \dodoi{10.1016/S0032-0633(01)00060-5}

\bibitem[{{Telloni} {et~al.}(2015){Telloni}, {Bruno}, \& {Trenchi}}]{Telloni15}
{Telloni}, D., {Bruno}, R., \& {Trenchi}, L. 2015, \apj, 805, 46,
  \dodoi{10.1088/0004-637X/805/1/46}

\bibitem[{{Velli} {et~al.}(1989){Velli}, {Grappin}, \& {Mangeney}}]{Velli89}
{Velli}, M., {Grappin}, R., \& {Mangeney}, A. 1989, \prl, 63, 1807,
  \dodoi{10.1103/PhysRevLett.63.1807}

\bibitem[{{Weck} {et~al.}(2015){Weck}, {Schaffner}, {Brown}, \&
  {Wicks}}]{Weck15}
{Weck}, P.~J., {Schaffner}, D.~A., {Brown}, M.~R., \& {Wicks}, R.~T. 2015,
  \pre, 91, 023101, \dodoi{10.1103/PhysRevE.91.023101}

\bibitem[{{Weygand} \& {Kivelson}(2019)}]{Weygand19}
{Weygand}, J.~M., \& {Kivelson}, M.~G. 2019, \apj, 872, 59,
  \dodoi{10.3847/1538-4357/aafda4}

\end{thebibliography}




\end{document}